\documentclass[runningheads]{llncs}
\usepackage{graphicx}
\graphicspath{ {images/} }
\usepackage{subfig}
\usepackage{gensymb}
\begin{document}
\title{Foveated Haptic Gaze}
%
%
\author{Bijan Fakhri\inst{1} \and Troy McDaniel\inst{1} \and Heni Ben Amor\inst{2}
 \and Hemanth Venkateswara\inst{1} 
 \and Abhik Chowdhury\inst{1} \and Sethuraman Panchanathan\inst{1}}

%
\authorrunning{B. Fakhri et al.}
%

\institute{The Center for Cognitive Ubiquitous Computing \\
\url{cubic.asu.edu} \\
\and
The Interactive Robotics Lab \\
\url{interactive-robotics.engineering.asu.edu} \\ 
Arizona State University, Tempe AZ 85281, USA\\}

\maketitle              
%
\begin{abstract}
	As digital worlds become ubiquitous via video games, simulations, virtual and augmented reality, people with disabilities who cannot access those worlds are becoming increasingly disenfranchised. 
	More often than not the design of these environments focuses on vision, making them inaccessible in whole or in part to people with visual impairments. 
	Accessible games and visual aids have been developed but their lack of prevalence or unintuitive interfaces make them impractical for daily use. 
	To address this gap, we present Foveated Haptic Gaze, a method for conveying visual information via haptics that is intuitive and designed for interacting with real-time 3-dimensional environments. 
	To validate our approach we developed a prototype of the system along with a simplified first-person shooter game.
	Lastly we present encouraging user study results of both sighted and blind participants using our system to play the game with no visual feedback. 
\end{abstract}

\section{Introduction}
	Virtual worlds are becoming ubiquitous as digital technology permeates society, with augmented and virtual reality being the latest and most immersive manifestations. 
	Unfortunately, the visual domain is central to most virtual worlds, making them inaccessible to people with visual impairments. 
	People with visual impairments already face accessiblity hurdles when using technology but virtual worlds remain one of the most inaccessible mediums. 
	Two competing approaches exist to correct this dilemma.
	Designers of virtual worlds develop the environments with accessibility in mind in the first approach. 
	Secondly, accessiblity engineers develop tools to make existing virtual environments accessible. 
	While the first approach is gaining traction and public awareness, developers of virtual environments seem to have been excused of this responsibility as accessible virtual environments remain extraordinarily scarce. 
	The second approach has the potential to affect many existing environments. 
	An example of the effectiveness of the second approach is screenreader technology.
	Screenreaders made digital text and many of the invaluable capabilities of smartphones accessible to millions of people with visual impairments.
	
	In the same vein, we aim to develop transformative technologies to make virtual worlds as accessible to people with visual impairments as text-based ones. 
	Mimicing the characteristics of the human visual system that make it so well-suited for interacting with 3-dimensional environments, we developed ``Foveated Haptic Gaze'', an intuitive method for exploring visual environments with the sense of touch. 
	``Foveated Haptic Gaze'' makes use of an attentional mechanism similar to foveated vision that allows users to focus on objects while simultaneously allowing for peripheral awareness.
	This combination gives users the ability to explore an environment in detail while maintaining broader situational awareness, making ``Foveated Haptic Gaze'' one of the only vision-to-haptic interfaces flexible enough to generalize to the real world.
	
	To validate our approach, we developed a first-person shooter game based on Doom, a working prototype of the Foveated Haptic Gaze system, and performed a user study with both individuals that are sighted and individuals with visual impairments. 
	Seeking to develop an approach that is useful to people with limited or no sighted priors, our user study measured the in-game performance of both populations to understand the effects sighted priors have on our approach.
	Additionally, we sought to understand any nuances of non-sighted human-computer interaction for 3D visual environments that could inform future approaches. 
	
	\begin{figure}
		\centering
		\subfloat[][]{\includegraphics[height=1.2in]{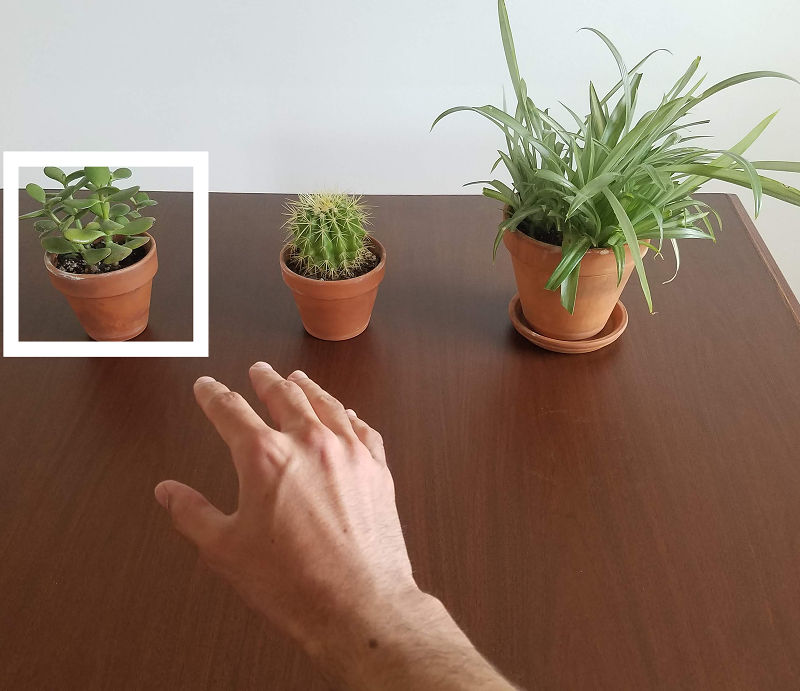}\label{fig:box1}}
		\qquad
		\subfloat[][]{\includegraphics[height=1.2in]{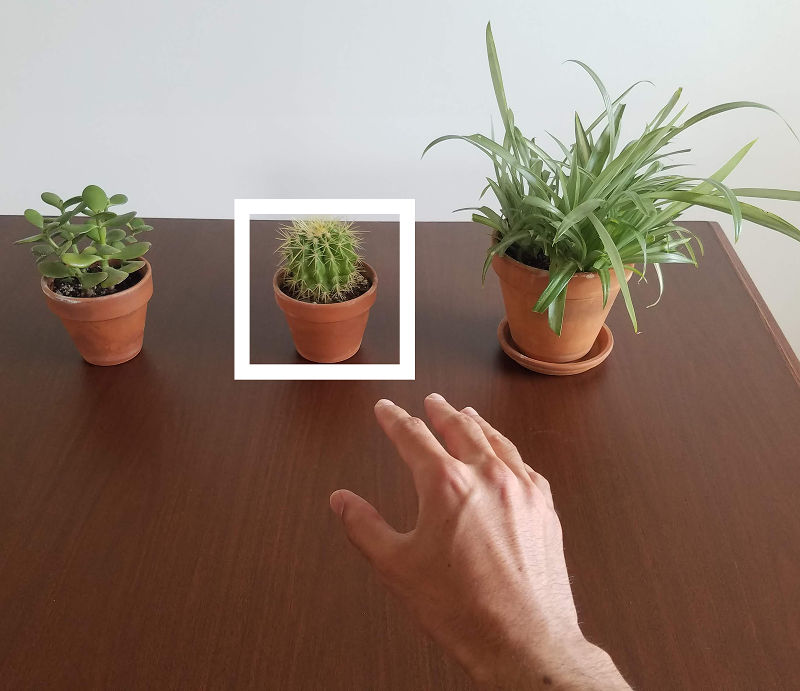}\label{fig:box2}}
		\qquad
		\subfloat[][]{\includegraphics[height=1.2in]{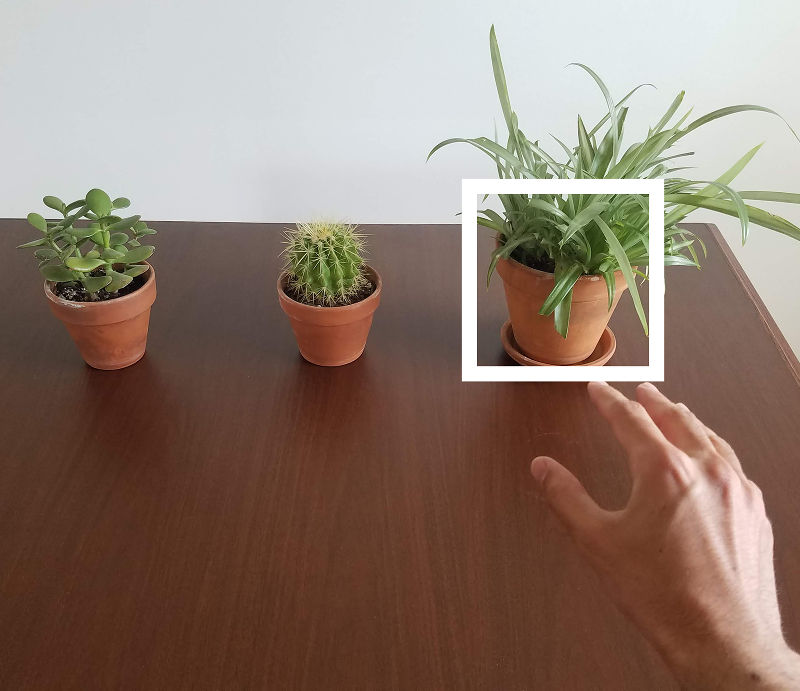}\label{fig:box3}}
		\caption{User's hand position determines where they are gazing: \protect\subref{fig:box1} Gazing at leftmost plant  \protect\subref{fig:box2} Gazing at middle plant \protect\subref{fig:box3} Gazing at rightmost plant}
	\end{figure}\label{fig:boxes}
	
\section{Related Works}
	\subsection{Non-visual Games}
		Accessiblity in games is becoming more and more popular. 
		It is no longer uncommon to find color-blind friendly settings in games as well as subtitles and other accessibility features. 
		An example of this in the context of virtual reality is SeeingVR, a suite of VR tools for making VR environments more accessible to people with low vision \cite{Zhao2019}.  
		Truly non-visual video games though have yet to become mainstream. 
		While non-visual video games are few and far between, there does exist a small collection. 
		Some of the first non-visual video games were developed for academic purposes such as the Audio-based Environment Simulator (AbES) games.
		AbES is a software suite designed to improve real world navigation skills for people with blindness \cite{Connors2013}. 
		AudioDOOM and AudioZelda \cite{SANCHEZ2009} \cite{Mirsky2009} were developed using AbES. 
		AudioDOOM  is one such AbES game that discritized a 3D environment into voxels that a user's avatar (and other entities) can move through via adjacent voxels. 
		Users could interact with entities such as monsters by fighting them when in the same voxel, although no aiming mechanics were involved. 
		After playing the game, children were asked to recreate the virtual environment using legos rendering promising results for the development of spatial awareness in the virtual world. 
		In AudioZelda, users navigate a college campus collecting items to develop familiarity with the campus' layout. 
		A more recent serious game for developing spatial skills is called Hungry Cat \cite{Chai2019}. 
		Researchers designed audio cues users could use for interacting with 3-dimensional maps. 
		The learned layouts were confirmed using physical representations similar to the validation of learned maps in AudioDOOM. 
		Similarly, researchers in \cite{Fakhri2019} developed a completely non-visual 2D game using a haptic chair interface where the objective was to move your avatar to a goal position. 
		This environment though did not rely on audio, users could feel the position of their avatar and the goal on their back using the haptic chair as they navigated the environment. \\
		
		Non-visual games for commercial and entertainment purposes also exist. 
		One of the most popular audio-only video games was called Papa Sangre and its successor Papa Sangre 2 \cite{PapaSangre2011}. 
		The game was an immersive adventure game based on 3D audio whereby the user navigates solely by listening to their surroundings (you can hear sleeping monsters whom you must not wake by stepping on) and tapping the screen to walk through the world. 
		Sadly the game is no longer available as of this writing. 
		A more recent iPhone app game is an audio-only ``Endless Runner'' game called FEER \cite{Meyer2018} \cite{Rego2018} whereby a user runs across a platform dodging enemies and collecting power-ups. 
		FEER received high praise from the American Foundation for the Blind (AFB) \cite{AFB_FEER}.
		Timecrest: The Door, is a story game where one's character has the power to control time and their decisions alter the course of the story \cite{TimeCrestAV}  \cite{SneakyCrab2017}. 	
		A Blind Legend is an action-adventure game where you fight with a sword and, similar to Papa Sangre, uses a 3-dimensional sound engine to create realistic and immersive soundscapes\cite{Dowino2019}. 
		The game has been well received by the community receiving 16,000 ratings with an average of 4.4 stars as of this writing.  
		All of these environments were designed to be used without a visual representations from the ground up. 
		Inversely, there have been a few efforts to make visual environments accessible via assistive technology. 

	\subsection{General Tools for Interacting with Visual Environments}
		Most famously, Dr. Bach-y-Rita's work on Sensory Substitution showed that after extensive training individuals with blindness were able to interact with visual stimuli via other sensory channels. 
		The first example of this was the Tactile-to-Vision Sensory Substitution (TVSS) system, a dental chair outfitted with actuators a seated person could feel on their back  \cite{Bach-Y-Rita1969}  \cite{White1970}.
		The next generation of these machines used electro-tactile stimulation via a tongue-display-unit \cite{Sampaio2001} to make the device more portable although slightly more intrusive.
		Less complex, consumer grade devices have also been developed for less serious applications. 
		Researchers Wall and Brewster compared two such devices with traditional raised-paper diagrams to assess their effectiveness in conveying visual information. 
		The devices compared were the VTPlayer, a computer mouse augmented with braille-like pins for providing cutaneous haptic feedback and the WingMan Force Feedback mouse which provides kinesthetic haptic feedback.  
		Researchers found raised paper to be the most effective while the WingMan Force Feedback Mouse and VTPlayer mouse were followed in effectiveness in that order. 
		Wall and Brewster hypothesized that the combination of kinesthetic and cutaneous haptic cues of the raised paper made it most effective in conveying visual information \cite{Wall2006b}. 
		
		One of the most exciting developments in this field is the emergence of Computer Vision methods that are useful for interacting with visual environments. 
		The social media giant Facebook already performs automatic image captioning on uploaded images, updating their alt-text dynamically \cite{Facebook16}.
		The explicitly ``assistive'' apps  Google Lookout and Microsoft Seeing AI give users audio descriptions of scenes captured on a user's phone that are intended to aid in understanding their surroundings \cite{Lookout2018} \cite{Microsoft18}. 
		Lookout alerts the user of the presence of some objects and their relative location while Seeing AI has a more comprehensive toolchest, sporting document reading and illumination descriptions capabilities.  
		While these methods are incredibly encouraging due to the richness of information they provide, their not yet real-time interfaces do not promote intuitive interaction with the visual world.
		They provide descriptions and summarizations of visual content, which while impressive and useful in some contexts, hinder a user's agency to explore the visual world deliberately. 

		One such device that encourages active exploration is the Auditory Night Sight \cite{Twardon2013}. 
		Researchers developed a system whereby eye-tracking technology was employed to control what portion of a depth map was relayed via audio to a user's ears (tone depicted depth values). 
		The concept of directing attention via the eyes is compelling: sighted individuals do this intuitively with gaze. 
		But solely providing point-depth cues does little for scene understanding and peripheral awareness.
		To be truly useful for interacting with rich visual environments, a device must provide real-time feedback, be intuitive and exploratory in nature, and grant the user agency and focus without sacrificing the expansive situational awareness made possible by natural peripheral vision. 
		
		\begin{figure}
			\centering
			\subfloat[][]{\includegraphics[height=1.05in]{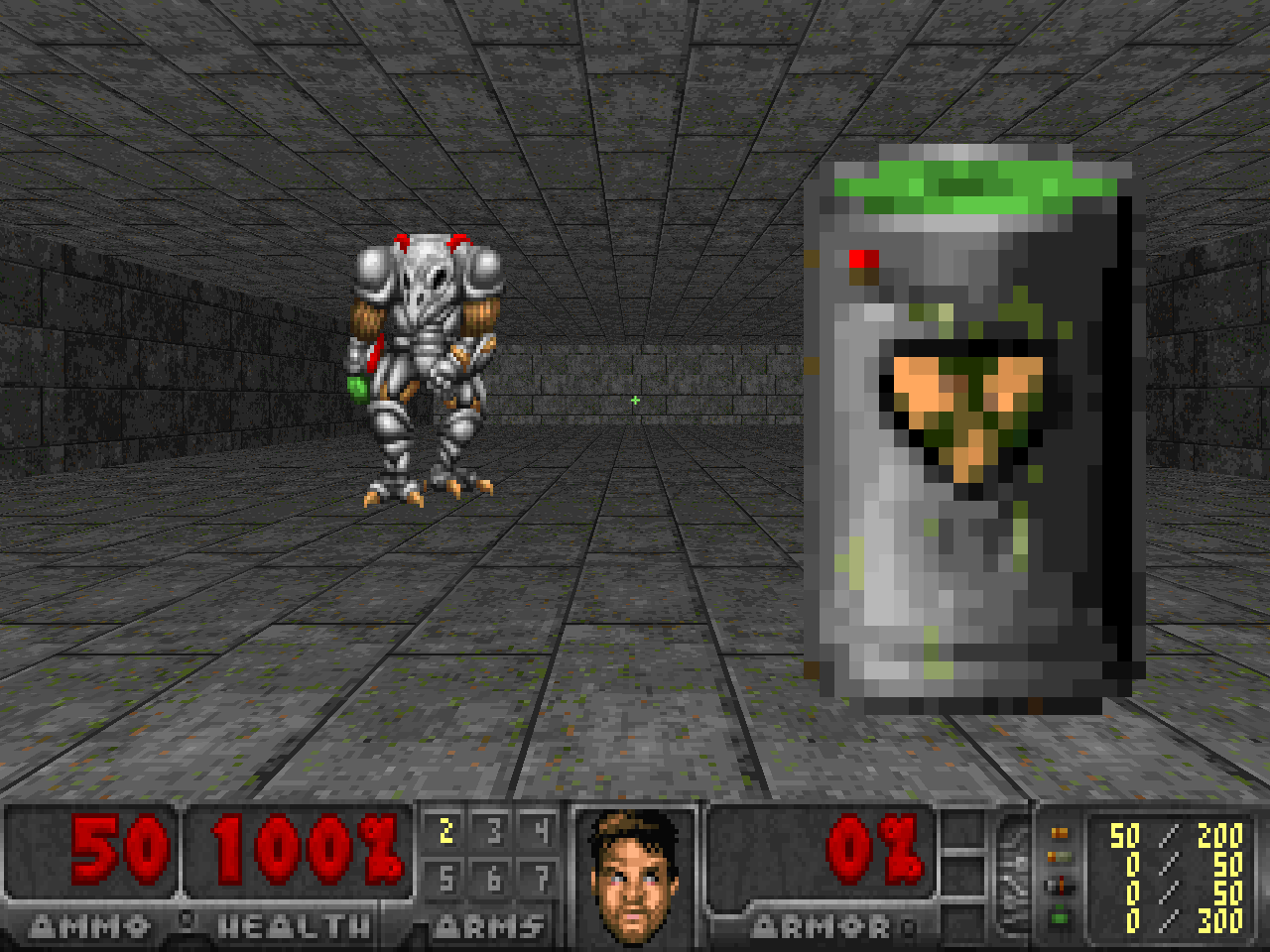}\label{fig:test_room_og}}
			\qquad
			\subfloat[][]{\includegraphics[height=1.05in]{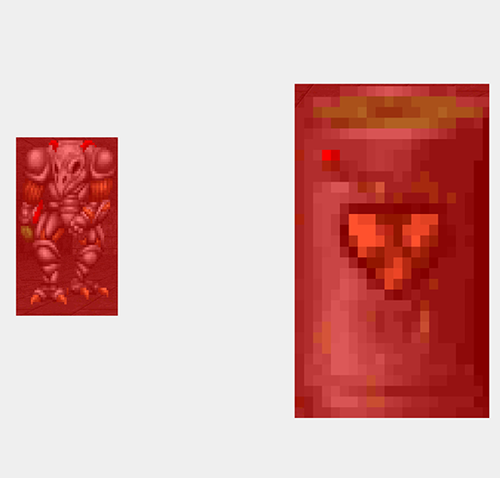}\label{fig:test_room_labels}}
			\qquad
			\subfloat[][]{\includegraphics[height=1.05in]{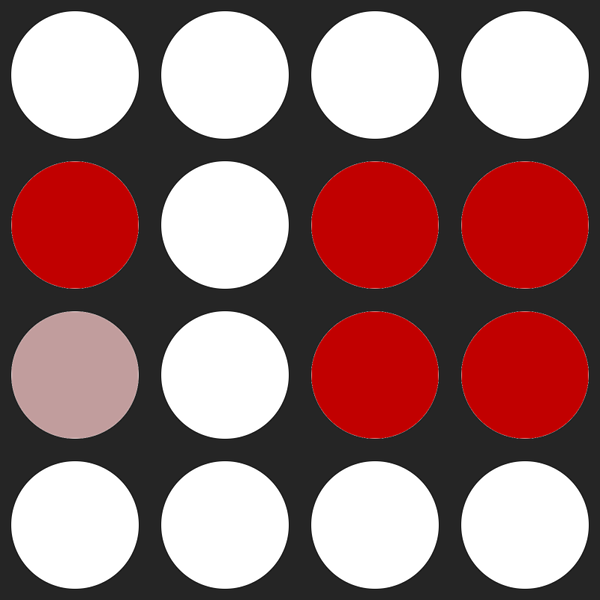}\label{fig:test_room_depth}}
			\caption{\protect\subref{fig:test_room_og} Original image of room  \protect\subref{fig:test_room_labels} Objects of interest highlighted \protect\subref{fig:test_room_depth} Corresponding motor array activations}
		\end{figure}\label{fig:test_room}

\section{Method}
	Human gaze is characterized by aligning the optical axis of the eye to whatever in the visual field one is interested in. 
	The optical axis also happens to be aligned with the fovea, an area of the retina featuring the highest density of photosentitive receptors \cite{NeuralScience}. 
	Gazing is thus directing one's visual attention by aligning the most acute portion of the retina with whatever is of interest. 
	The rest of the retina is responsible for peripheral vision, enabling a wide (up to $220\degree$ horizontally) spatial awareness in direct spatial relation to one's focus \cite{Szinte2012}. 
	Thus the human visual system has the capacity for high resolution as well as  expansive field-of-view thanks in part to foveated vision. 
	
	\subsection{Foveated Haptic Gaze}
\begin{figure}
			\centering
			\subfloat[][]{\includegraphics[height=1.05in]{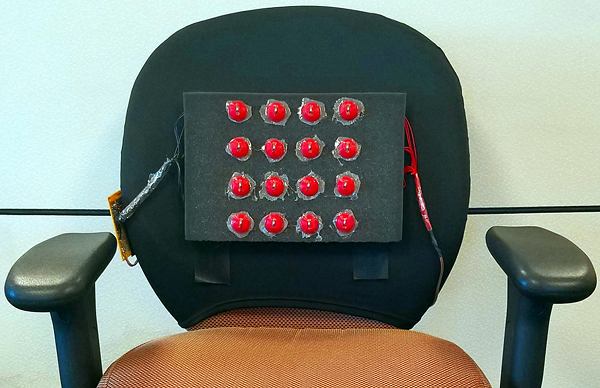}\label{fig:chair}}
			\qquad
			\subfloat[][]{\includegraphics[height=1.05in]{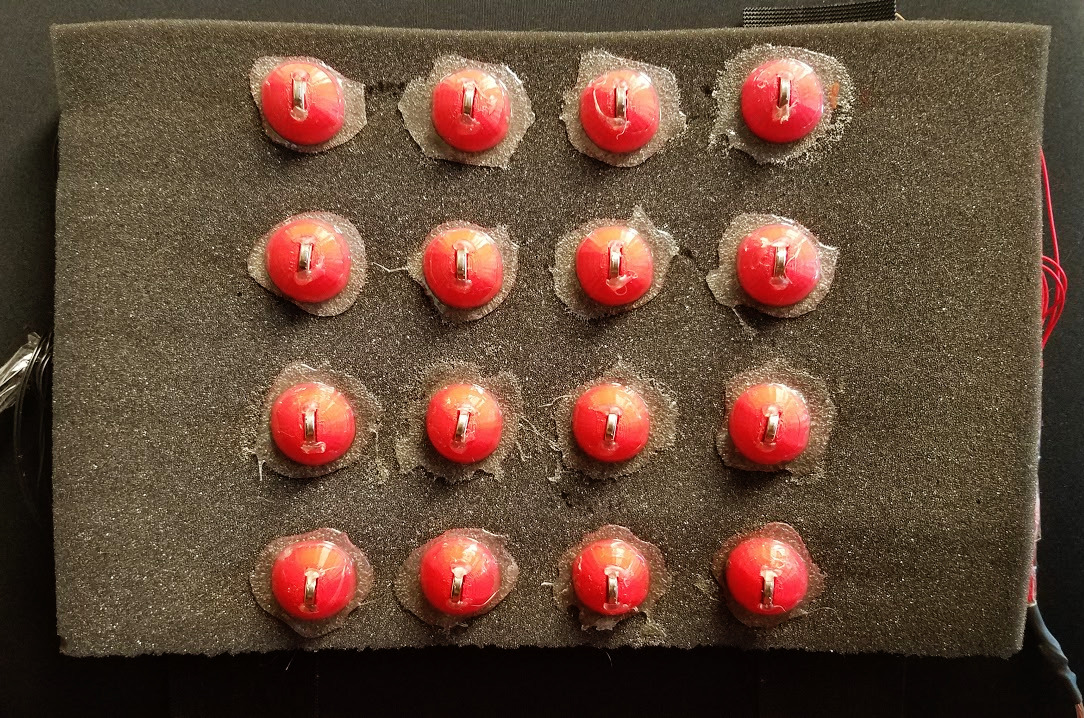}\label{fig:motor_array}}
			\caption{\protect\subref{fig:chair} Haptic display on office chair   \protect\subref{fig:test_room_labels}  Closeup of motor array}
		\end{figure}

		We borrow the concept of foveated vision to develop a biologically inspired haptic implementation called Foveated Haptic Gaze (FHG). 
		In the same way sighted individuals gaze with their eyes by pointing their foveas at objects of interest, using our system individuals with visual impairments can gaze in a visual environment by pointing their hand at objects of interest (an illustration can be seen in figure \ref{fig:boxes}). 
		The user wears a purpose built haptic glove (shown in figure \ref{fig:glove}) and when they point their hand at an object, details of the object are haptically conveyed via the glove equiped with vibration motors on the finger tips. 
		This provides an analog to the high-resolution fovea, while a back-mounted haptic display (shown in figure \ref{fig:chair}) \cite{Fakhri2019} endows the user with peripheral awareness (Haptic Peripheral Vision) of their entire field-of-view. 
		The system thus partitions a user's experience into two channels: one for high-fidelity and one for wide field of view. 
		The back display alerts the user to the presence and coarse location of objects (obstacles, doors, persons, etc) while pointing a hand towards these objects provides the user with finer details of the object, such as the object's identity (e.g. ``door'', ``person'', etc). 
		To integrate these two systems so that a user can relate the position of their haptic gaze with their haptic peripheral vision, the system displays the position of their gaze with respect to their field-of-view on the back display. 
		Practically, a user feels on their back where objects are and where their gaze currently is, moving their hand to align these indicators is essentially gazing at the object. 
		This is akin to noticing an object in your periphery then gazing at it for more details. 
		To illustrate the effectiveness of our approach we created a gaming environment with which participants can interact with rich 3D spatial situations.
		
	\subsection{Gaming Environment}\label{sec:doom_environment}
\begin{figure}
			\centering
			\includegraphics[height=2.0in]{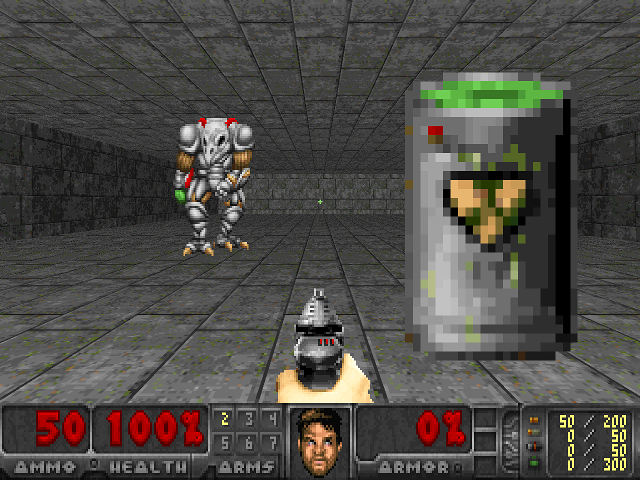}
			\caption{Doom Environment featuring a ``Hell Knight'' monster on the left and explosive barrel on the right.}
			\label{fig:doom}
		\end{figure}

		The First-Person Shooter (FPS) genre of video games was a natural choice for testing the system's efficacy because FPSs offer a realistic simulation of the first-person experience as well as mechanics like aiming and shooting that require keen visuospatial awareness to play effectively. 
		The game DOOM is one of the most iconic and modded FPS games in existence, making it our choice for developing experimental environments using the ViZDoom platform.
		ViZDoom \cite{Kempka2017} enabled us to develop visually rich, low-overhead, and responsive DOOM environments for use in our experiments. 
		A system that can empower users to effectively play a game like DOOM has the best chances of generalizing to real-world interactive visual environments. 
		Figure \ref{fig:doom} shows an image of the environment from the first person perspective. 

\begin{figure}
			\centering
			\includegraphics[width=\textwidth]{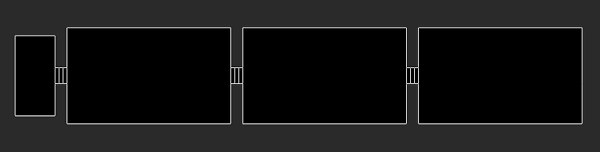}
			\caption{Bird's eye view of the (abridged) game map used in the study. The full map consisted of 10 interconnected rooms.}
			\label{fig:map}
		\end{figure}

		We designed a level consisting of 10 connected rooms.
		The player runs through the rooms encountering monsters and explosive barrels (shown in figure \ref{fig:doom}). 
		Figure \ref{fig:map} shows a top-down view of the rooms: there are 11 monsters and 5 explosive barrels randomly positioned in the rooms, with more monsters/barrels occuring in later rooms. 
		The objective is to shoot as many monsters as possible while not shooting the explosive barrels. 
		The player's score is the difference between the number of monsters killed and the number of explosive barrels shot: $score = monsters-barrels$. 
		A user willl feel the presence and position of monsters or barrels in their field of view on their back via the haptic display. 
		To ascertain whether the objects are monsters or barrels, the user must gaze over the object with their hand. 
		
	\subsection{System Design}
\begin{figure}
			\centering
			\includegraphics[width=\textwidth]{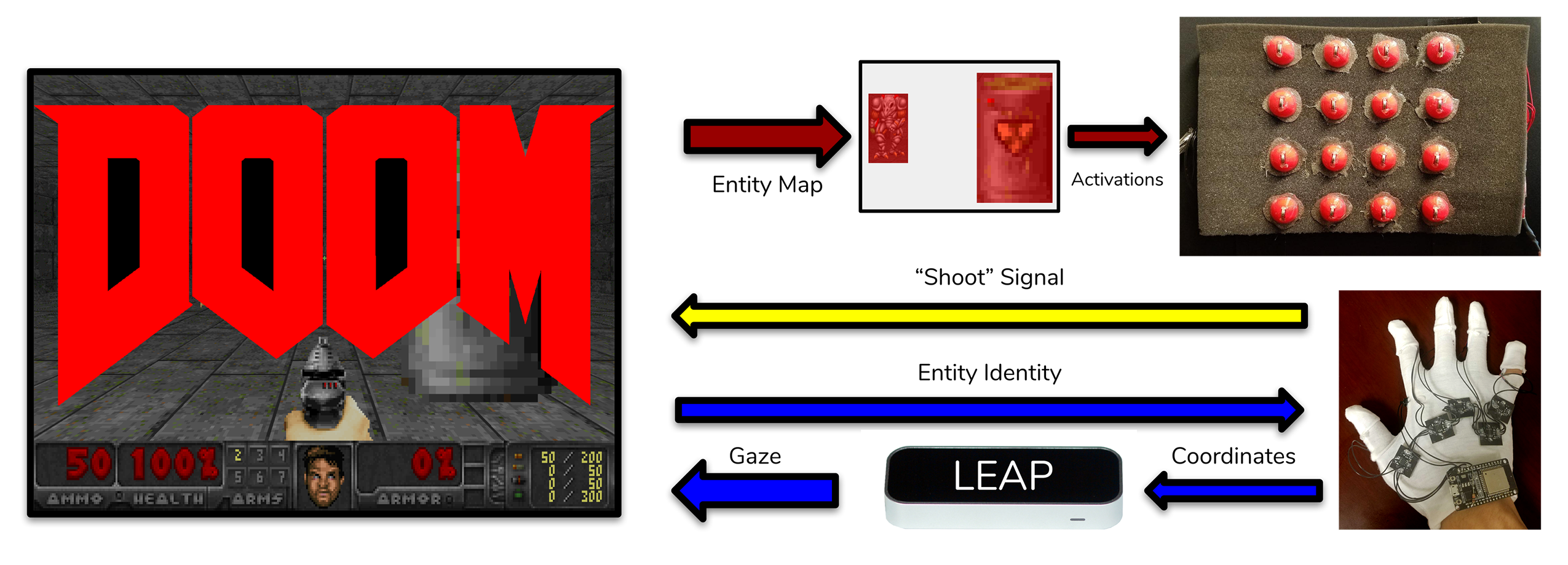}
			\caption{Diagram of experimental setup. The ViZDoom game engine sends an entity map to the haptic display (red) to be felt by the user. The hand's movements are tracked by the Leap Motion Controller and its position is converted to gaze coordinates on the avatar's field of view. If the gaze intersects with any entities their identity is sent to the glove (blue). If the user presses the trigger button on the glove a shoot signal is sent to the ViZDoom environment and the avatar fires in the direction the user is gazing. }
			\label{fig:wholesystem}
		\end{figure}

		A user wears a glove equiped with a button and vibration motors on the finger tips (shown in figure \ref{fig:glove}). 
		The vibration motors convey information about what the user is gazing at, and in the case of our hallway game, reveal to the user whether they are gazing at a monster or a barrel. 
		The user's hand position is tracked with a Leap Motion Controller, and the 3D coordinates of the hand are mapped onto the field of view of the player's avatar. 
		We extract from the ViZDoom environment the location of obstacles in the avatar's field of view and map this information as well as the user's gaze position onto the haptic display on the user's back. 
		A diagram of the whole system can be seen in figure \ref{fig:wholesystem}. 

		\begin{figure}
			\centering
			\subfloat[][]{\includegraphics[width=2.05in]{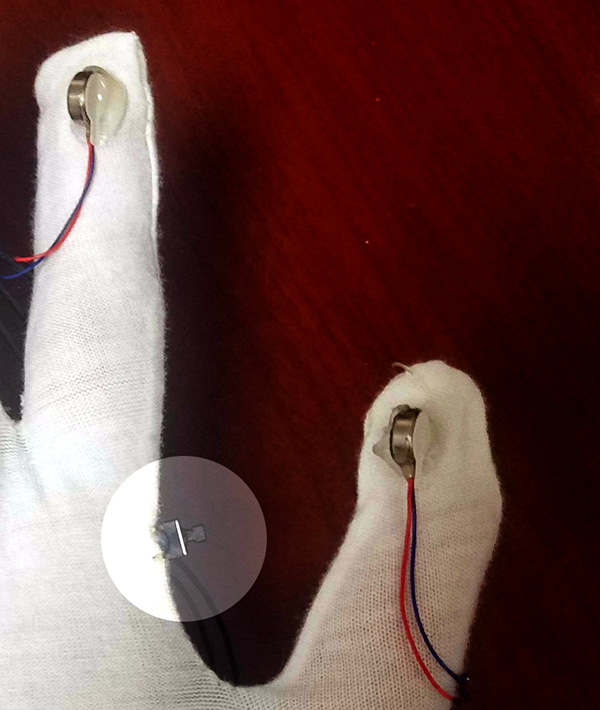}\label{fig:glove_front}}
			\qquad
			\subfloat[][]{\includegraphics[width=2.05in]{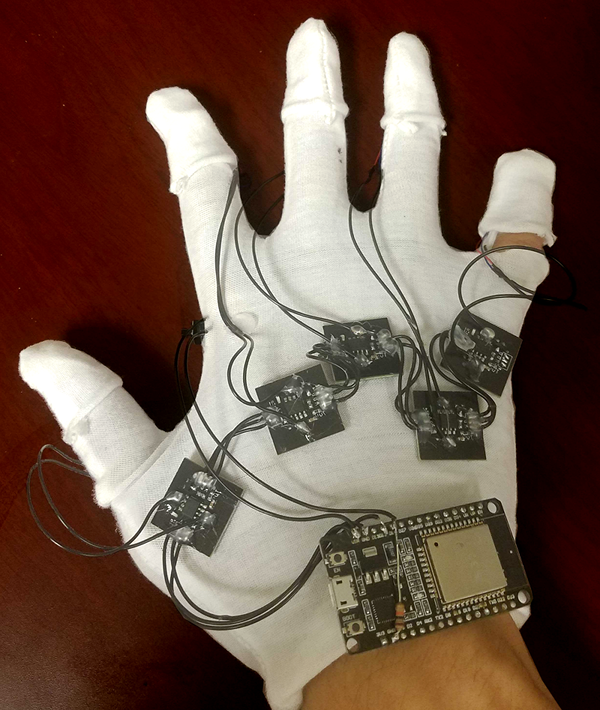}\label{fig:glove_back}}
			\caption{\protect\subref{fig:glove_front} Pancake motors and button (highlighted) on haptic glove \protect\subref{fig:glove_back}  Image of the back of the glove with motor driving hardware shown}
		\end{figure}\label{fig:glove}
		
	\subsection{Experimental Design}	
		Five participants with visual impairments and ten sighted participants were recruited for the user study. 
		At the beginning of the study, participants were acquainted with the hardware they would be using: haptic display (chair), haptic glove, and Leap Motion Controller.
		Participants were then introduced to the concept of FHG by performing an introductory exercise that activated the Leap Motion Controller and haptic display only. 
		The participant's hand was tracked and displayed on their back using the haptic display and participants were encouraged to acquaint themselves with the limits of their field of view.
		The purpose of this exercise was to illustrate the mechanics of the gazing mechanism e.g. moving one's hand to the left moved their gaze to the left on their back. 
		Next they were introduced to the concept of Haptic Peripheral Vision.
		
		Users' avatars were placed in a room in the ViZDoom environment populated by one monster and one explosive barrel on either size of their field of view. 
		The haptic chair relayed the locations of the monster and barrel to them by pulsating on their backs (see figure \ref{fig:test_room_depth}). 
		The location of their gaze was also conveyed by the haptic display via a solid vibration; consequently users learned to gaze towards the objects in the room by aligning the gaze vibrations with the pulsating ``entity'' vibrations on their back. 
		Upon placing their gaze over one of the entities (monster or barrel), the identity of the entity was conveyed to the user via the glove's vibration motors in a coded manner.
		Users were instructed to discriminate a ``monster pattern'' and ``barrel'' pattern. 
		After exploring the room by gazing over the entities, participants were instructed to shoot both entities. 
		When the barrel is shot it explodes and creates a load explosion sound while shooting the monster results in a triumphant ``winning'' sound.
		These are the only audio cues in the whole game other than rhythmic game music. 
		
		After this explaination, participants were asked if they were comfortable with the interface and objective and were given the chance to enter the demo room once again, after which the experiment began. 
		Participants entered the hallway game environment described in section \ref{sec:doom_environment} and illustrated in figure \ref{fig:map} to score as many points as possible. 
		Participants were asked to play 7 games (each taking about 1.5 minutes to complete) and their performance as well as auxiliary metrics (shots fired, hits, misses) were recorded during their gameplay. 
		
\section{Results}	
\begin{figure}
		\centering
		\subfloat[][]{\includegraphics[width=2.2in]{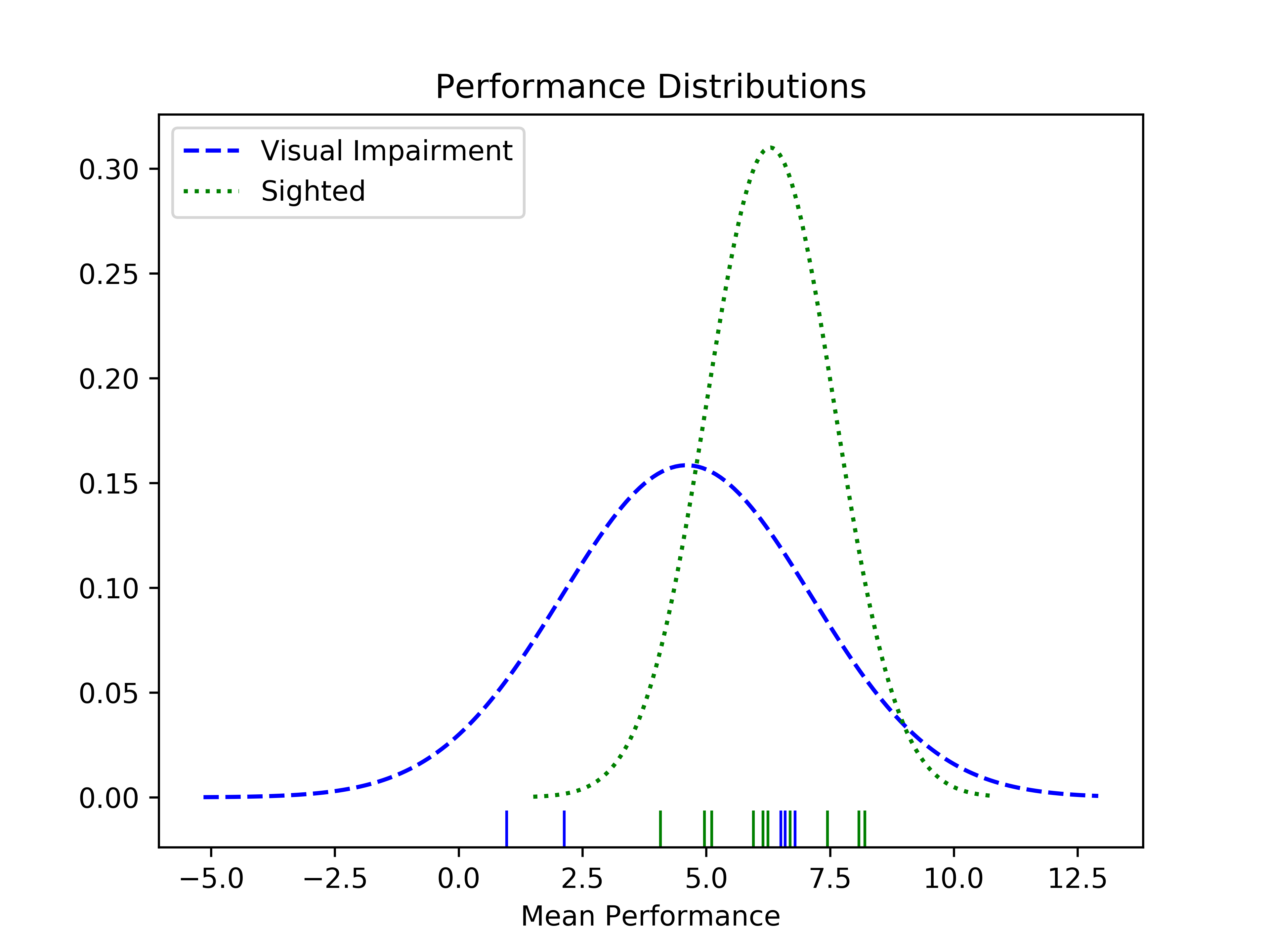}\label{fig:perfnorms}}
		\qquad
		\subfloat[][]{\includegraphics[width=2.2in]{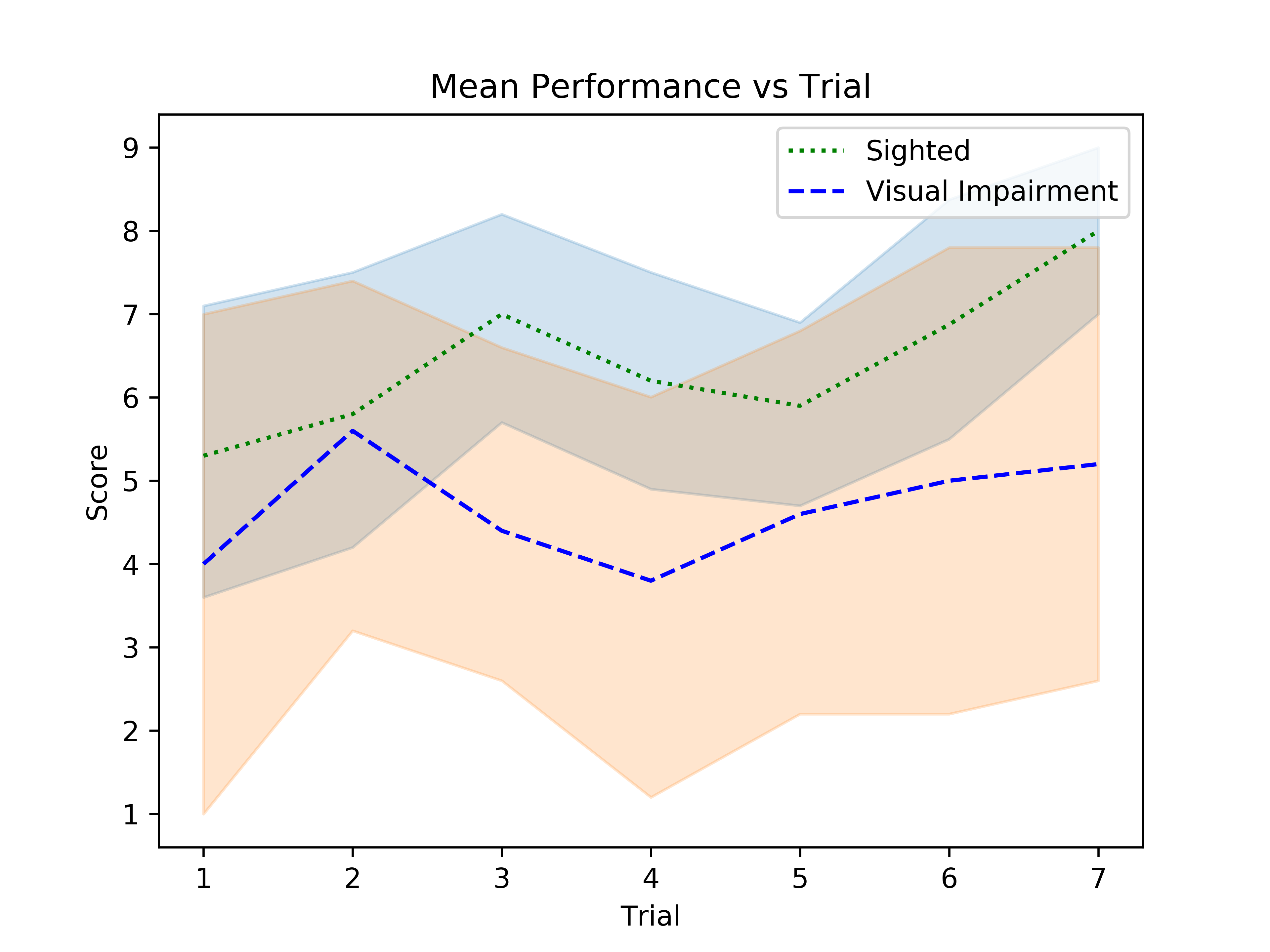}\label{fig:perf_over_time}}
		\caption{\protect\subref{fig:perfnorms} Normal distribution fit to the performance of both participant populations averaged over all trials \protect\subref{fig:perf_over_time} Performance over time averaged over participants}
	\end{figure}

	To assess playability as well as any differences in usability between sighted and users with visual impairments, we measured a player's score throughout every game played. 
	On average, sighted users obtained higher scores although the majority of users with visual impairments also clustered towards the center of the sighted performance distribution shown in figure \ref{fig:perfnorms}. 
	Both populations saw an initial increase in performance although sighted individuals maintained an upward trajectory slightly longer while participants with visual impairments leveled off sooner. 
	Figure \ref{fig:perf_over_time} illustrates their performance over time. 
	The theoretical maximum score is 11 as there are 11 monsters to destroy, although their positioning often makes them difficult to destroy due to their brief visibility. 
	
\begin{figure}
		\centering
		\subfloat[][]{\includegraphics[height=1.25in]{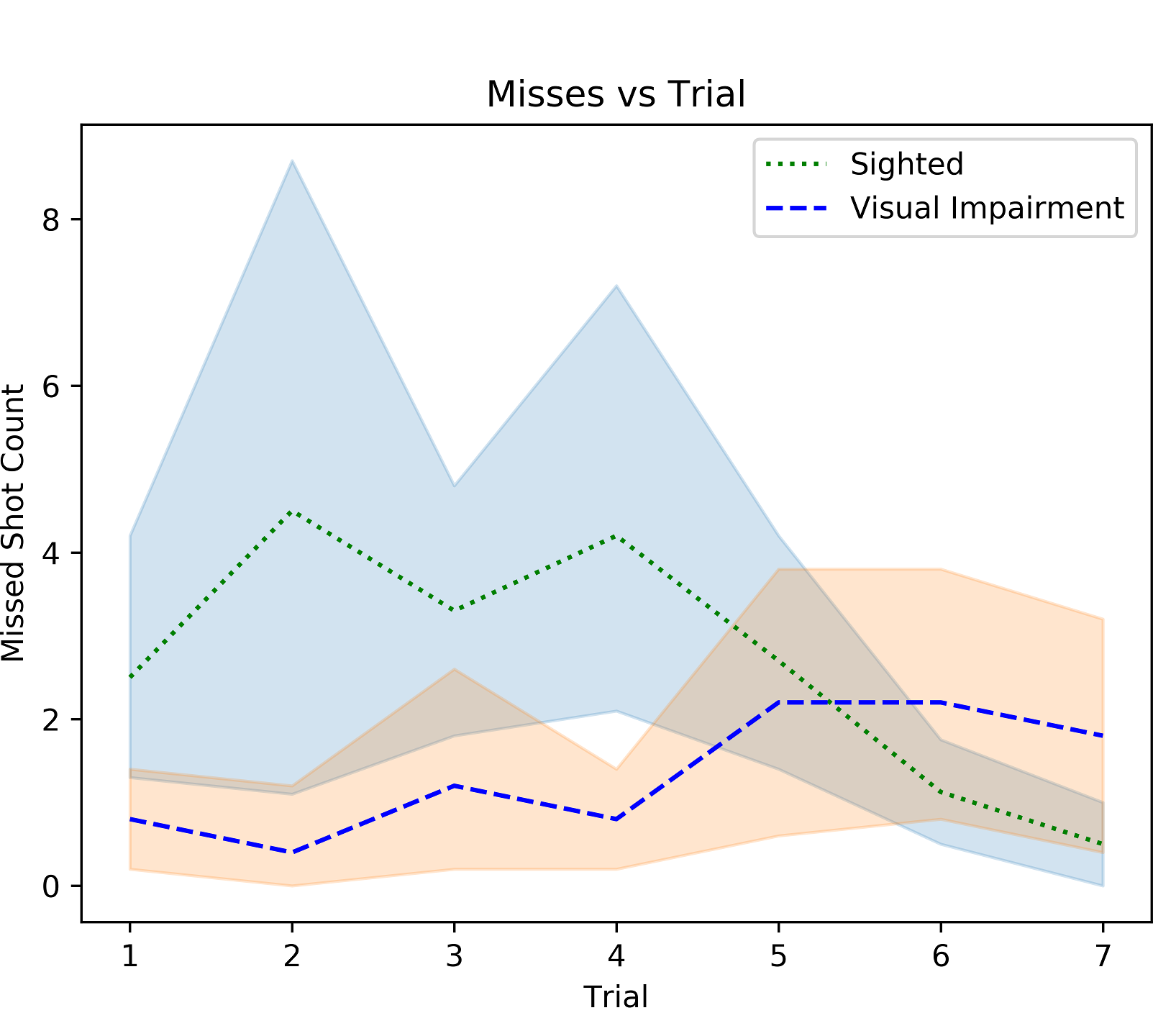}\label{fig:misses_over_time}}
		\qquad
		\subfloat[][]{\includegraphics[height=1.25in]{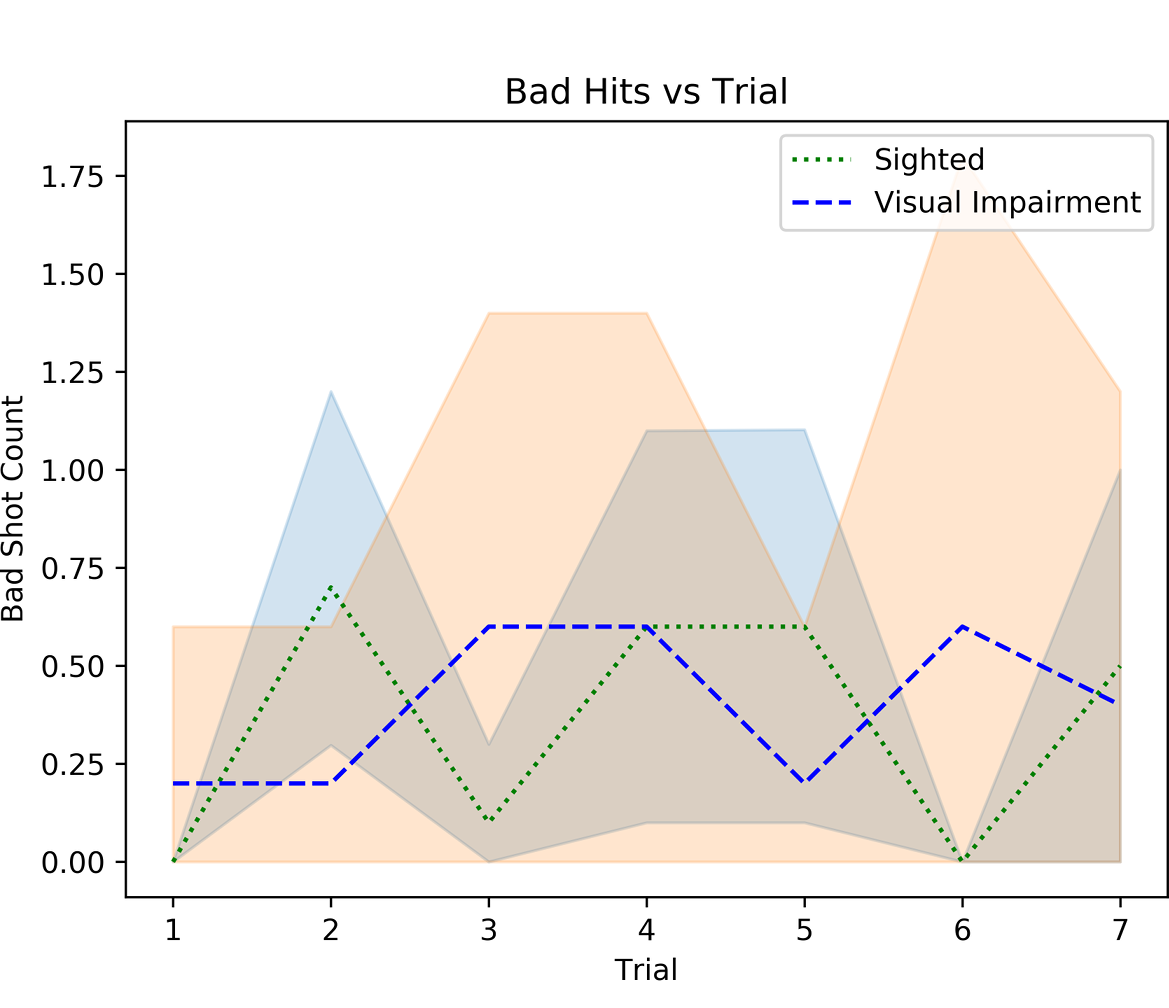}\label{fig:badhits_over_time}}
		\qquad
		\subfloat[][]{\includegraphics[height=1.25in]{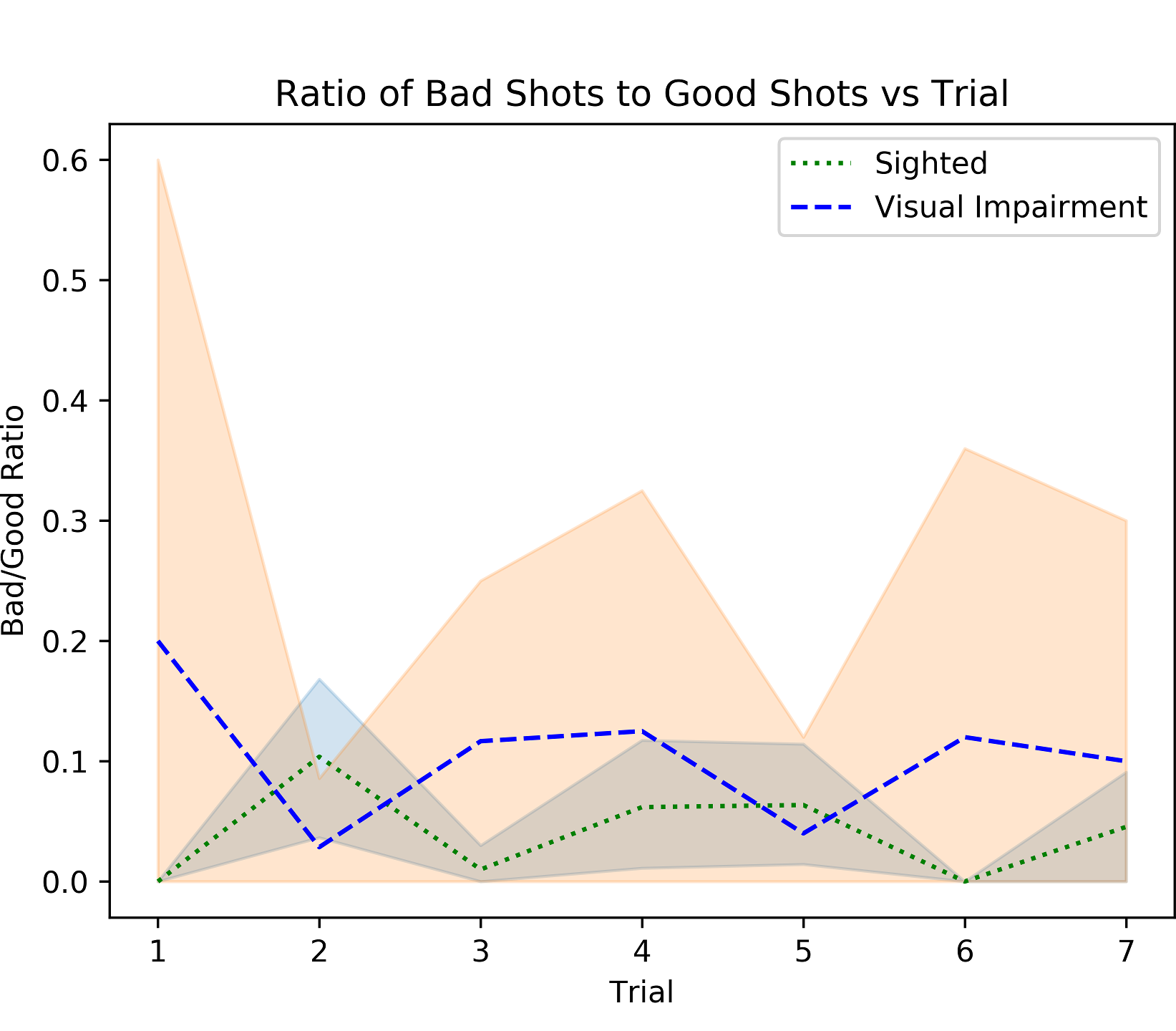}\label{fig:gobot}}
		\caption{\protect\subref{fig:accuracy_over_time} Misses per trial \protect\subref{fig:badhits_over_time} Shots that hit a barrel per trial: mistakes made by participants by shooting an entity they were instructed not to shoot \protect\subref{fig:gobot} Ratio of enemies killed to explosive barrels (mistakes) over trial. There is a large variance in performance initially for participants with visual impairments that quickly dwindles as the participants learn from their mistakes.}
	\end{figure}

	To assess a player's ability to make decisions on-the-fly, they were instructed to avoid shooting explosive barrels, as it would negatively impact one's score. 
	These mistakes as well as good hits and complete misses were recorded on a per-game basis. 
	Players overal made few mistakes, many averaging below one mistake per game (figure \ref{fig:badhits_over_time}), indicating that the glove feedback was clear and intuitive: as a ratio of mistakes to good shots (monsters killed), most players stayed below $1/10$. 

\begin{figure}
		\centering
		\subfloat[][]{\includegraphics[width=2.2in]{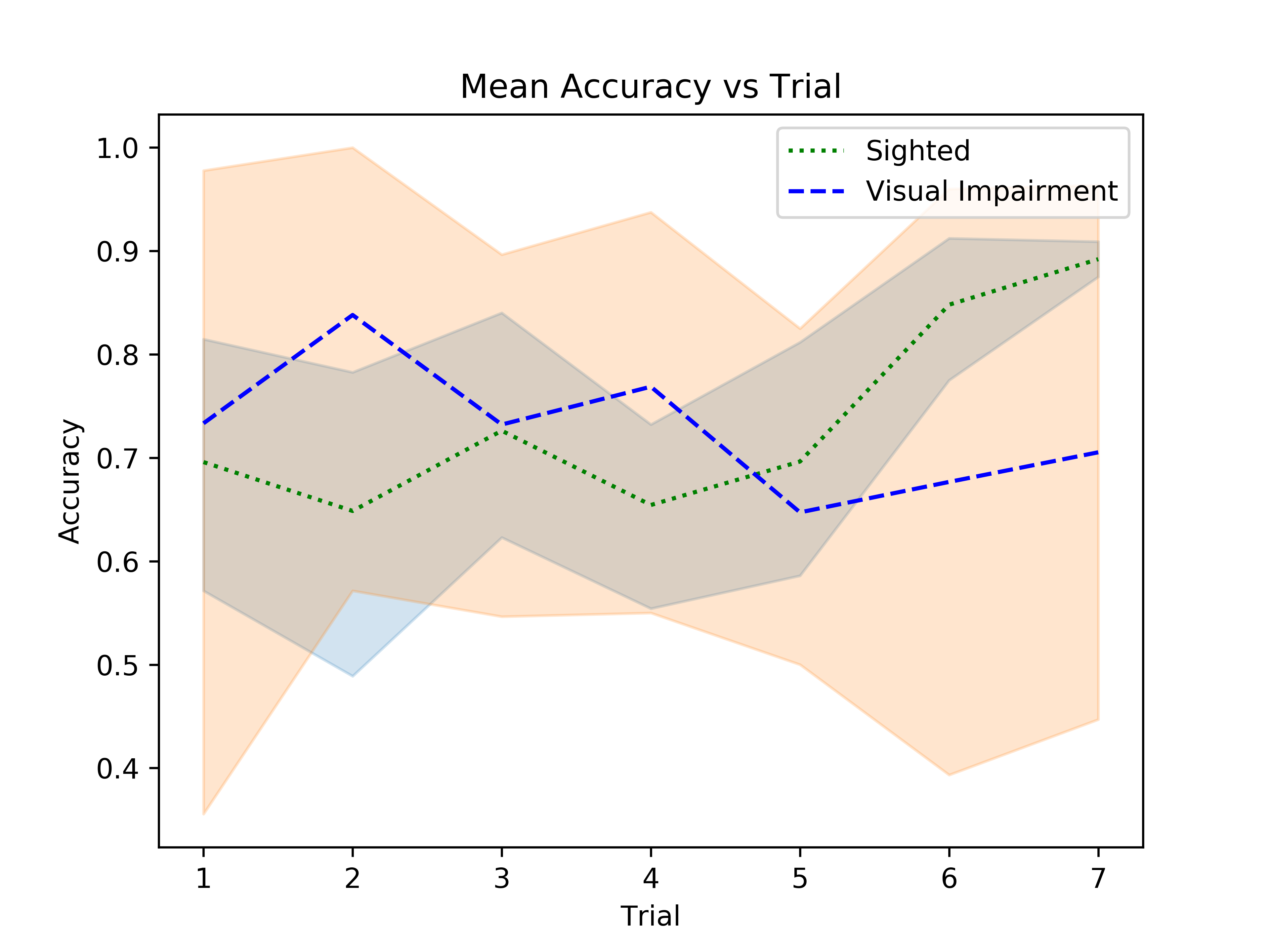}\label{fig:accuracy_over_time}}
		\qquad
		\subfloat[][]{\includegraphics[width=2.2in]{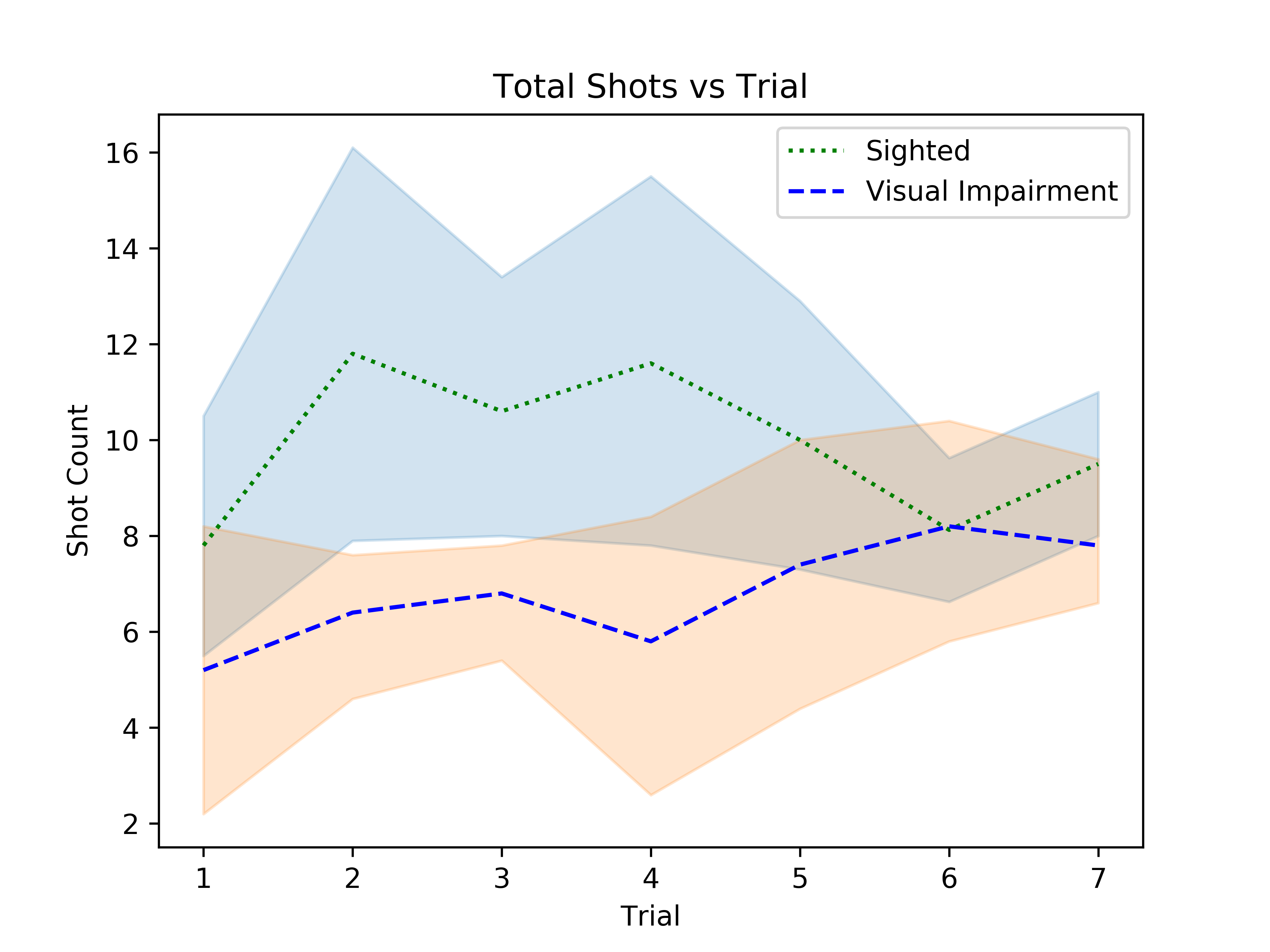}\label{fig:shots_over_time}}
		\caption{\protect\subref{fig:accuracy_over_time} Accuracy over time (hits / shots taken) averaged over all participants in each populaiton \protect\subref{fig:shots_over_time} Total shots taken per trial, averaged over all participants in each population}
	\end{figure}

	Participants with visual impairments initially missed less than sighted participants, trending upwards throughout the trials eventually ending slightly higher than sighted participants (figure \ref{fig:misses_over_time}) . 
	Inversely, sighted participants missed more often from games 1 through 5, but during the last two games ended with slightly fewer average misses. 
	These trends imply that participants with visual impairments tended to approach the game more cautiously than sighted individuals, becoming more comfortable as games went on while their sighted counterparts were more cavalier to begin with and reigned in their enthusiasm as the games progressed. 
	This is supported by the total shot counts per trial figure, plotted in figure \ref{fig:shots_over_time}, where it can be seen that sighted participants initially took many more shots than those with visual impairments. 

	Both sets of participants performed similarly with regards to accuracy (hits over total shots taken per game) as illustrated in figure \ref{fig:accuracy_over_time}. 
	Players achieved an accuracy between $70-80\%$ during the first 5 games, indicating that the aiming and gazing mechanics of the system were usable for real-time interactions. 
	Interestingly, sighted players' accuracy rose to touch $90\%$ during the last two games, in tandem with their dip in misses (figure \ref{fig:misses_over_time}). 

\section{Conclusion and Future Work}
	Results from our user study indicated the playability of the Doom game was maintained without vision as most participants were able to achieve respectable performance metrics and accuracies. 
	This was supported by positive subjective user feedback with regards to the system design. 
	Differences in performance between test groups were small, boding well for our approach having only slight sighted usability bias. 
	A more extensive analysis is required to rule out a sighted performance bias and may inform design decisions to make the approach even more intuitive to people without vision. 
	The results also indicate that individuals with visual impairments approached the game more cautiously, becoming less cautious over time while sighted participants approached the game with less caution and became slightly more cautious over time. 
	Consequently, future approaches may benefit from designs that encourage confidence inspiring exploration. 
	Furthermore, the presentation of peripheral vision information (Haptic Peripheral Vision) can likely be improved.
	The accuracy assessments indicate that foveated gaze feedback worked well, while destroying all 11 monsters remained difficult for both populations, as brief appearances of monsters were sometimes missed. 
	A higher resolution haptic back display or one with wider back coverage such as the HaptWrap \cite{Duarte19} may mitigate this by providing more salient peripheral awareness feedback.

	For future work, we plan to take a more practical implementation, modifying our system for real-world use.
	We seek to couple the Foveated Haptic Gaze technology with computer vision techniques and wearable haptic displays such as the HaptWrap to generalize to complex real-world environments. 
	
\section{Awknowledgements}
	The authors would like to thank Arizona State University and the National Science Foundation for their funding support. This material is partially based upon work supported by the National Science Foundation under Grant No. 1828010.


\bibliographystyle{splncs04}
\bibliography{../../BibTex_Mendeley/Main-ICSM19.bib}

\end{document}